\documentclass{ere04}

\usepackage{graphicx}    

\def\beq{\begin{equation}}
\def\eeq{\end{equation}}

\def\bea{\begin{eqnarray}}
\def\eea{\end{eqnarray}}
\def\ba{\begin{array}}                  
\def\ea{\end{array}}

\newcommand{\pa}{\partial}

\begin{document}
\title*{On the gauge invariance of the non-Abelian Chern-Simons action for D-branes}
\author{J. Adam \inst{1}
        \and J. Gheerardyn \inst{2}
        \and B. Janssen \inst{3}
        \and Y. Lozano \inst{4}}
\institute{Katholieke Univ. Leuven, Celestijnenlaan 200 D, Leuven, Belgium 
%
\and Universt\`a di Torino and INFN Sezione di Torino, Via Giuria 1, 10125 Torino, Italy
%
\and Universidad de Granada and CAFPE, Campus Fuente Nueva s/n, 18071 Granada, Spain 
%
\and Universidad de Oviedo, Avda. Calvo Sotelo 18, 33007 Oviedo
}

\maketitle

We present an elegant method to prove the invariance of the Chern-Simons part of 
the non-Abelian action for N coinciding D-branes under the R-R and NS-NS gauge 
transformations, by carefully defining what is meant by a background gauge 
transformation in the non-Abelian world volume action. We study as well the 
invariance under massive gauge transformations of the massive Type IIA supergravity 
and show that no massive dielectric couplings are necessary to achieve this 
invariance.\footnote{Talk given by B.J. at the XXVII Spanish Relativity Meeting in 
Madrid, September 2004.}


\vspace{1cm}

It is well known by now that the physics of a set of $N$ coincident D$p$-branes 
can be very different from the physics of $N$ parallel but separated D$p$-branes.  
Witten showed \cite{Witten} that in the former case a number of new massless 
states appear that can  be arranged in representations of a  $U(N)$ gauge group. 
In particular, the $N$ Born-Infeld vectors form a single $U(N)$ Yang-Mills 
vector $V_a$ and the transverse scalars, arranged in $N \times N$ matrices 
$X^i$, become non-Abelian matrices transforming in the adjoint representation of 
$U(N)$, where the $I$-th eigenvalue of the matrix $X^i$ has the interpretation of 
the position of the $I$-th D-brane in the direction $x^i$. 

The new physics associated to these extra massless string states has to be encoded 
in the world volume effective action, which now should be written in terms of the 
matrix valued fields $V_a$ and $X^i$. Determining the exact form of the
Born-Infeld action is a highly non-trivial problem, to
which the solution is still not clear (see for instance \cite{Tseytlin2}).
A lot of progress has been made however over the last few years in the understanding
of the structure of the non-Abelian Chern-Simons (or Wess-Zumino) action.

The first generalisation of the Chern-Simons term
to the $U(N)$ case was proposed in \cite{Douglas}:
\bea
S_{{\rm D}p}\ =\ T_p \int P[C] \ \mbox{Tr} \{ e^{\cal F} \}
            \ = \ T_p \int \sum_n P[C_{p-2n+1}]\  \mbox{Tr} \{ {\cal F}^n \}.
\label{GHTac}
\eea
Here the trace is taken over the Yang-Mills indices of the $N$-dimensional 
representation of $U(N)$ and  $P[\Omega]$ denotes the pullback of the background 
field $\Omega$ to the world volume of the D-brane. The world volume field $\cal F$ 
is given by ${\cal F} = F + P[B]$, where 
$F = 2 \partial V + i [V, V]$ is the non-Abelian field strength 
of the Born-Infeld vector and $B$ the NS-NS two-form.

The invariance of this action under the gauge transformations of the background
NS-NS and R-R fields was further investigated in \cite{GHT}, where it was shown
as well that in order to be invariant under the massive gauge transformations of 
massive Type IIA supergravity \cite{Romans, BRGPT}, extra $m$-dependent terms 
were needed in the action. These extra terms were also obtained from the (massive) 
T-duality relations \cite{BRGPT, BR} between the different D-brane actions, 
generalising to the non-Abelian case the Abelian calculation of \cite{BR}.

Nowadays we know, however, that the Chern-Simons action for coincident D-branes 
presented in \cite{GHT} is not the complete story. On the one hand, in the 
non-Abelian case the background fields in (\ref{GHTac}) must be functionals of 
the matrix-valued coordinates $X^i$ \cite{Douglas2}. Explicit calculations of 
string scattering amplitudes \cite{GM} suggest that this dependence is given 
by a non-Abelian Taylor expansion
\beq
C_{\mu\nu} (x^a, X^i) = \sum_n {{\frac{1}{n!}}} \partial_{k_1} ... \partial_{k_n}
                     C_{\mu\nu} (x^a, x^i){\mid}_{x^i = 0} \
                      X^{k_1} ... \ X^{k_n}.
\label{taylorexp}
\eeq
On the other hand, in order to have invariance under $U(N)$ gauge transformations 
the pullbacks of the background fields into the world volume have to be defined in 
terms of $U(N)$ covariant derivatives $D_a X^\mu = \partial_a X^\mu + i[V_a, X^\mu]$,
rather than partial derivatives \cite{Dorn, Hull}. For instance\footnote{From now 
on instead of working in the static gauge we will write everything in a 
``diffeomorphism invariant'' way, with the understanding that $U(N)$ covariant 
derivatives reduce to ordinary ones for $X^\mu$ lying in the world volume of the 
D-branes.},
\beq
P[C_2] = C_{\mu\nu} \ D_{[a}X^\mu D_{b]}X^\nu.
\label{pullback}
\eeq
This, together with the symmetrised trace prescription \cite{Tseytlin}, that we 
will denote by curly brackets $\{..\}$, assures the invariance of the action under 
$U(N)$ gauge transformations
\beq
\delta V_a = D_a \chi,  \hspace{2cm}
\delta X^i = i [\chi, X^i].
\eeq
Finally, the most important modification to the action (\ref{GHTac}) is the presence
of new dielectric couplings to higher order background field potentials, arising as a 
consequence of T-duality in non-Abelian actions \cite{Myers, TvR}. It was found that
the full T-duality invariant form of the Chern-Simons action
is given by:
\bea
S_{{\rm D}p}\ =\ T_p \int \Bigl\{ 
P[e^{({\mbox i}_X {\mbox i}_X)} (C e^B) ] \  e^F \Bigr\},
\label{myersactie}
\eea
where $({\mbox i}_X C)_{\mu_1\dots \mu_n}$ denotes the interior product
$X^{\mu_0} C_{\mu_0\dots \mu_n}$.

One should note however that the presence of $U(N)$ covariant pullbacks has 
consequences on the invariance under gauge transformations of the background 
fields. Let us look for example at the variation $\delta C_{\mu\nu} = 
\partial_{[\mu} \Lambda_{\nu]}$ of the term given in (\ref{pullback}). Naively 
filling in the variation in the pullback yields:
\beq
\delta \{ P[C_2]\}  = \{ P[\partial \Lambda_1] \}
          = \{ \partial_\mu\Lambda_\nu \ D_{[a} X^\mu D_{b]}X^\nu \}.
          \label{derivtot}
\eeq
In the Abelian limit this gauge variation is a total derivative, such that the
$\Lambda_1$ gauge invariance is assured in the D1-brane Chern-Simons action. 
In the non-Abelian case however the variation is not a total derivative
such that not even D1-branes with topologically trivial world volumes are 
described by a gauge invariant action. The same goes for the non-Abelian couplings 
present in  (\ref{myersactie}): a pullback of a variational parameter of the form 
$\{P[ ({\mbox i}_X {\mbox i}_X) \partial \Lambda_{n+1} ]\}$ is by no means a 
total derivative. 

It is clear from these examples that the question of how to perform background 
gauge transformations in the action (\ref{myersactie}) is far from obvious. As the 
gauge transformations themselves are given by supergravity and the form of the 
action is derived in various compatible ways, we can not change these (too much). 
The only way therefore to construct a gauge invariant action is to change the way 
these transformations are implemented in the action.

\vspace{.4cm}

Let us first concentrate on the simplest case of the monopole terms in 
(\ref{myersactie}), setting for now, all dielectric couplings to zero.
In order to have an action invariant under the background gauge transformations, 
we need to fulfill three conditions. First, it must be possible to write the 
variation as a total derivative, secondly, the variation has to be a scalar 
under $U(N)$ gauge transformations and finally, it has to reduce to the known case
in the Abelian limit. Therefore we define the variation of the pullback of a R-R field 
$C_p$ under the background gauge transformation 
$\delta C_{p} = \partial \Lambda_{p-1}$ as \cite{AGJL}:
\bea
\delta  P[C_{p}] \Omega \ \equiv \    D P[\Lambda_{p-1}] \Omega
                         \  =  \  D_{[a_1|} (\Lambda_{\mu_2 ... \mu_{p}}
                                      D_{|a_2}X^{\mu_2}... D_{a_p]}X^{\mu_p} )\Omega,
\label{defvar}
\eea
where $\Omega$ is any combination of world volume or pullbacked background fields 
and where it is understood that all $U(N)$-valued objects appear symmetrised (though 
not in a trace). In particular for the simplest case with $\Omega= 1$ we find that 
\bea
\delta  P[C_p]
   &=&  \   D_{[a_1|} (\Lambda_{\mu_2 ... \mu_{p}}
                                  D_{|a_2}X^{\mu_2}... D_{a_p]}X^{\mu_p} )\\
   &=&   P[\partial \Lambda_{p-1}] \
           + \ {\textstyle{\frac{i}{2}}} (p-1)
\ \Lambda_{\mu_1...\mu_{p-1}} [F_{[a_1 a_2}, X^{\mu_2}]
                    D_{a_3}X^{\mu_3}... D_{a_{p}]}X^{\mu_{p}}. \nonumber
\eea
With this definition we see that the variation is not just the pullback of the 
gauge pa\-ra\-meter, but contains as well a non-Abelian correction term proportional 
to $[F, X]$, since the covariant derivative  $D_{a_1}$ not only acts on the 
background gauge parameter $\Lambda_{p-1}$, but also on the covariant derivatives 
in the pullback. For the Abelian case, the correction term disappears and we recover 
the well-known gauge transformation for Abelian D-brane actions. Furthermore once 
we consider terms in the action and trace over all $U(N)$ indices in the 
symmetrised trace prescription the variation is in fact a total derivative:
\beq
\delta \{ P[C_{p}] \} = \{ D P[\Lambda_{p-1}]\}
                      = \partial \{ P[\Lambda_{p-1}]\}.
\label{deltaC}
\eeq
In general for the background gauge transformations
$\delta C_p = \sum_n \partial \Lambda_{p-2n -1}B^n  -m \Sigma B^{(p-1)/2}$,
we define the pullbacks in the action to vary as
\beq
\delta  P[C_p]\Omega = \sum_n DP[ \Lambda_{p-2n -1}] P[B^n]\Omega 
         \ - \ m P[\Sigma B^{(p-1)/2}]\Omega.
\label{RRabels}
\eeq
Similarly the non-Abelian version of the NS-NS gauge transformation 
$\delta B = \partial \Sigma$ is given by 
\beq
\delta P[B]\Omega = 2 D P[\Sigma] \Omega,   \hspace{1cm}
\delta V=-P[\Sigma].
\label{NSabels} 
\eeq
Note that the Born-Infeld field transforms as well, such that the non-Abelian 
field strength ${\cal F} = F + P[B]$ is an invariant quantity, as should be expected 
from the Abelian case.

With these definitions, the computation of the gauge transformations of the action
\beq
{\cal L} =\Bigl\{ \sum_n P[C_{p-2n+1}] {\cal F}^n + m\  \omega_{2n+1} \Bigr \}
\eeq
is straightforward, since it formally reduces to the Abelian case. Note that an extra
Chern-Simons-like term \cite{GHT}
\bea
\omega_{2n+1} = \sum_k \  V (\partial V)^{n-k} [V, V]^k
\label{omega}
\eea
had to be added to the action of the even D-branes, in order to assure 
the invariance under the massive gauge transformations. These terms  are constructed 
in such a way that they transform under the Yang-Mills gauge transformations as a 
total derivative, and under the $\Sigma$ transformations as
\beq
\delta \omega_{2n+1} = -  \Sigma F^n,
\eeq
and thus cancel the massive gauge transformation of the R-R background fields.

\vspace{.4cm}
So far we have rederived the results of \cite{GHT} on the gauge invariance of 
non-Abelian Chern-Simons actions, taking into account explicitly the $U(N)$ 
covariant pullbacks and the fact that the background fields are functionals 
of the non-Abelian coordinates $X^\mu$. As we have seen this forces a precise 
definition for what we mean by gauge variation of a non-Abelian pullback. A 
consistency check of our definitions (\ref{RRabels})-(\ref{NSabels}) is that 
the variation of the pullback of a R-R $p$-form should be T-dual to the 
variation of the pullback of a R-R $(p-1)$-form field. We will now check this 
and see that in this manner we can find a natural way to also prove the gauge 
invariance of the dielectric terms.


To show this let us define a R-R field ${\tilde C}_p$, being related to $C_p$ 
via a gauge transformation
${\tilde C}_P= C_p +  \partial \Lambda_{p-1}$.
We then have on the one hand by definition (\ref{defvar}) that
\beq
P[{\tilde C_p}] = P[C_p] + DP[\Lambda_{p-1}]
\eeq
while on the other hand we know from \cite{Myers} that by applying T-duality on 
${\tilde C}_p$ we get (for simplicity we truncate for now to the 
``diagonal approximation''  $g_{\hat\mu x} = B_{\hat\mu\hat\nu} = 0$)
\bea
 P[{\tilde C}_p]    
&& \rightarrow  P[{\tilde C}_{p-1}]  \ +\ i P[({\mbox i}_X {\mbox i}_X) {\tilde C}_{p+1}]  
\nonumber \\[.3cm]
&&=   P[{C}_{p-1}]  \ +\ i P[({\mbox i}_X {\mbox i}_X) {C}_{p+1}]  
    \  + \  DP[\Lambda_{p-2}] +   \ i DP[({\mbox i}_X {\mbox i}_X) {\Lambda}_{p}] 
\eea
where we used that ${\tilde C}_{p-1}$ and ${\tilde C}_{p+1}$ are related to, respectively, 
$C_{p-1}$ and $C_{p+1}$ by the same type of background gauge transformation that 
relates ${\tilde C}_p$ to $C_p$. We then find that the pullback of the gauge parameter 
transforms under T-duality as
\beq
 D P[\Lambda_{p}]  \rightarrow
 \ DP[ \Lambda_{p-1}]
\ +\  \ i \ D P[({\mbox i}_X {\mbox i}_X) \Lambda_{p+1}] .
\eeq
In other words, the variation of the pullback of a R-R $p$-form potential
goes under T-duality to the
variation of the pullback of a R-R $(p-1)$-form potential
plus the variation of the pullback of the first dielectric coupling term:
\beq
\delta  P[C_p]  \ \rightarrow \
\delta  P[C_{p-1}] \  + \ i \ \delta  P[ ({\mbox i}_X {\mbox i}_X) C_{p+1}] ,
\eeq
if we define:
\bea \delta P[ (i_Xi_X) C_{p+1}] 
&\equiv&  \partial  P[ ({\mbox i}_X {\mbox i}_X) \Lambda_{p}].
\eea
The derivation with the full T-duality rules (beyond the diagonal approximation) is straightforward
and not very enlightening, so we rather concentrate on the generalisation of the variation
(\ref{RRabels}) for dielectric couplings,
which can be derived in a similar way. Under general R-R gauge
transformations,  the dielectric terms vary as
\bea
\delta  P[({\mbox i}_X {\mbox i}_X) C_{p}] 
        &=& \sum_{n} \Bigl(
              D P[ ({\mbox i}_X {\mbox i}_X) \Lambda_{p-2n-1}] P[B^n]  
            +\  D P [{\mbox i}_X\Lambda_{p-2n-1}]
                                     P[({\mbox i}_X B)B^{n-1}]  \nonumber  \\
       && \hspace{-1cm}
            +\  D P [\Lambda_{p-2n-1}]
                                     P[({\mbox i}_X B)^2 B^{n-2}]  
            +\  D P [\Lambda_{p-2n-1}]
                                     P[({\mbox i}_X {\mbox i}_XB) B^{n-1}] \Bigr).
\label{Lambdadielec}
\eea
Note that the inclusion factor $({\mbox i}_X {\mbox i}_X)$ acts on the various background fields.
Similarly, under massive gauge transformations, the dielectric terms
transform as
\beq
\delta  P[({\mbox i}_X {\mbox i}_X ) C_{p}] = 
- \ m \ P[ ({\mbox i}_X {\mbox i}_X)(\Sigma B^{(p-1)/2}) ].
\label{mdielec}
\eeq

As an example let us now look at the gauge transformations of the non-Abelian action for
D6-branes, being the simplest non-trivial case in which both dielectric couplings and
massive gauge transformations are present. For this case, the non-Abelian Chern-Simons action  
can be written as
\bea
{\cal L}_{D6} \sim  \Bigl\{ \sum_{n} 
                         P\Bigl[ ({\mbox i}_X {\mbox i}_X) {\cal A}_{9-2n} \Bigr] F^n \Bigr\},
\label{nonAbaction}
\eea
where the $p$-forms ${\cal A}_{p}$ are defined as ${\cal A}_p= \sum_{k} C_{p-2k} B^k$.
It is obvious from the Abelian case that each ${\cal A}_{p}$ is invariant under the R-R 
and massive gauge transformations, such that the invariance of the action (\ref{nonAbaction})
under the transformations (\ref{RRabels}), (\ref{Lambdadielec}) and (\ref{mdielec}) is 
straightforward. It is also clear that besides the massive terms (\ref{omega}), 
introduced in \cite{GHT}, no other dielectric mass terms are needed to assure gauge 
invariance. This can also be confirmed by deriving the action by performing massive T-dualities
from the D9-brane action \cite{AGJL}.

The invariance under the NS-NS transformations (\ref{NSabels}) is however more subtle,
due to the fact that $({\mbox i}_X {\mbox i}_X)$ acts on $B$ but  not on $F$, 
so that they do not combine in an obvious way into the interior product of the gauge 
invariant field strength $\cal F$. In order to show the invariance under these 
transformations let us rewrite (\ref{nonAbaction}) as a function of the $C_p$, rather 
than ${\cal A}_{p}$, similar to the form of the action used in (\ref{GHTac}):
\bea
{\cal L}_{D6} &\sim&  \Bigl\{ \sum_{n} 
P\Bigl[ ({\mbox i}_X {\mbox i}_X) C_{9-2n} {\cal F}^n \
        +   \ ({\mbox i}_X   C_{9-2n}) ({\mbox i}_X B)  {\cal F}^{n-1} \nonumber \\
&& \hspace{2cm}
        + \ C_{9-2n} ({\mbox i}_X B)({\mbox i}_X B)  {\cal F}^{n-2} \
       + \   C_{9-2n} ({\mbox i}_X{\mbox i}_X B)  {\cal F}^{n-1}\
\Bigr] 
\Bigr\},
\label{nonAbaction2}
\eea 
Again here the inclusion terms  $({\mbox i}_X {\mbox i}_X)$ act both on $C$ as on $B$. Note 
that all the $B$ fields that are not acted upon by an inclusion term combine with the BI
field strength $F$ into the gauge invariant ${\cal F}$. However, the $B$'s contracted 
with one or more ${\mbox i}_X$ do not combine in a gauge invariant quantity and their 
variation can not be canceled by any other term in the action. The only field that also 
transforms under NS-NS transformations is the BI vector $V_a$, but for being a worldvolume 
fields it will never  appear contracted with ${\mbox i}_X$.

In \cite{AGJL} it was suggested that the variation of these terms is identically zero, 
due to the fact that translations in the transverse directions are isometries. Recall 
that the action for non-Abelian D$p$-branes is derived from the action for coincident 
D9-branes using T-duality \cite{Myers}, so that the directions in which the T-dualities 
are performed have to be isometric and hence the contractions of $\pa\Sigma$ with the 
transverse scalars must vanish, guaranteeing thus the gauge invariance of 
(\ref{nonAbaction2}). Furthermore it was suggested in \cite{AGJL} that since in the 
non-Abelian case there is no clear notion of general coordinate transformations (see 
for example \cite{DS}-\cite{BFLV}),  it is not clear how the resulting isometries can 
be removed.

However, there are now reasons to believe that the reasoning on \cite{AGJL} might not be 
completely correct, as phenomena such as the dielectric effect depend explicitly
on the coordinate dependence in the transverse directions. It has been suggested\footnote{We
thank Rob Myers for this comment.} that the variation of the ${\mbox i}_X B$ terms might be 
canceled by variations of other fields in the action that have not been taken into account 
yet. Indeed, it is not difficult to see that, after applying T-duality in a worldvolume 
direction $x$, the gauge variation of the $x$-component of $V$ leads to the following  
transformation of the new transverse scalar in the T-dualised action:
\beq
\delta X^x = \xi^x + i \Sigma_\mu [X^x, X^\mu], 
\label{varX}
\eeq
where $\xi^x$ is the T-dual of $\Sigma_\mu$ and plays (in the Abelian case) the role of a 
coordinate transformation, while the second terms suggest a kind of non-Abelian NS-NS gauge 
variation for the embedding scalars $X$.  

At this stage it is not clear what the interpretation of the variation (\ref{varX}) is 
(whether a coordinate transformation, or a gauge transformation) and whether it can be used 
to cancel the variations of the $({\mbox i}_X B)$ and $({\mbox i}_X{\mbox i}_X B)$ terms 
in (\ref{nonAbaction2}), but it does suggest that it might be helpful to use the 
well-known relation between NS-NS gauge transformations and coordinate transformations
through T-duality in order to learn more about the problem of general covariance of 
non-Abelian actions. We hope to report further progress in this direction soon \cite{AJ}.

\section*{Acknowledgments}
We wish to thank Mees de Roo, Martijn Eenink and Rob Myers for the useful discussions.
The work of J.A. is done as Aspirant F.W.O. She is also partially supported 
by the F.W.O.-Vlaanderen project G0193.00N and by the Belgian Federal Office 
for Scientific, Technical and Cultural Affairs through the Interuniversity 
Attraction Pole P5/27. 
The work of J.G. has been supported in part by the F.W.O.-Vlaanderen 
as postdoctoral researcher and by the Italian M.I.U.R. under the contract 
P.R.I.N. 2003023852. 
The work of B.J. is done as part of the 
program Ram\'on y Cajal of the M.E.C. (Spain). He was also partially supported by 
the M.E.C. under contract FIS 2004-06823 and by the Junta de Andaluc\'{\i}a group 
FQM 101.
The work of Y.L. has been partially supported by CICYT grants BFM2000-0357 and 
BFM2003-00313 (Spain).
J.A., J.G. and Y.L. are also partially
supported by the European Commission FP6 program MRTN-CT-2004-005104 in
which Y.L. is associated to Universidad Aut\'onoma de Madrid.


\end{document}